# Decoupling Josephson Coupling and Supercurrent Nonreciprocity in Twisted $NbSe_2/NbSe_2$ van der Waals Junctions

Seung-Gi Lee[1], Kwon-Neong Jung[1], Jae-Kuen Kim[3], Jae-Chun Jeon[2], Jiho Yoon[2], Stuart S. P. Parkin[2*], and Kun-Rok Jeon[1*]

[1]*Department of Physics, Chung-Ang University (CAU), 06974 Seoul, Republic of Korea*

[2]*Max Planck Institute of Microstructure Physics, Weinberg 2, 06120 Halle (Saale), Germany*

[3]*Department of Physics, University of Seoul, 02504 Seoul, Republic of Korea*

*To whom correspondence should be addressed: stuart.parkin@halle-mpi.mpg.de, jeonkunrok@gmail.com

**ABSTRACT**

**The microscopic origin of supercurrent nonreciprocity in van der Waals Josephson junctions remains under active debate, particularly regarding the role of twist-angle engineering in layered superconductors. Here, we investigate superconducting transport in twisted $NbSe_2/NbSe_2$ vertical Josephson junctions fabricated by dry transfer with controlled crystallographic alignment and chemically clean interfaces. High-resolution transmission electron microscopy is employed to directly determine the twist angle and assess interface quality. While the Josephson coupling strength exhibits a pronounced dependence on twist angle, with characteristic voltages maximized near crystallographically equivalent orientations and suppressed at intermediate angles, the supercurrent diode efficiency remains negligibly small and shows no systematic twist-angle dependence. In contrast, enhanced diode-like responses emerge only in weakly coupled junctions exhibiting interfacial disorder and reduced transparency. Deliberate**

**interface degradation further amplifies the apparent nonreciprocity, yielding diode efficiencies approaching 30% together with an irregular magnetic-field-strength dependence. These results establish a clear decoupling between Josephson coupling and supercurrent nonreciprocity in twisted $NbSe_2$/$NbSe_2$ junctions. Our findings identify interface disorder, rather than twist-angle-controlled momentum matching, as the dominant origin of the observed diode response and provide a critical benchmark for interpreting nonreciprocal superconducting transport in van der Waals Josephson devices.**

**KEYWORDS: Josephson coupling, supercurrent nonreciprocity, van der Waals superconductors, twisted $NbSe_2$/$NbSe_2$ junctions, interface disorder**

## 1. INTRODUCTION

Van der Waals (vdW) vertical Josephson junctions (JJs) provide a powerful platform for exploring superconducting transport in atomically engineered interfaces, enabling both device applications and the study of emergent quantum phenomena[1–8]. These structures, typically assembled from two-dimensional (2D) superconductors, offer clean and highly controllable interfaces in which electronic coupling can be tuned with high precision[9–11]. Among the available tuning parameters, the crystallographic twist angle $\theta_{twist}$ between stacked layers has emerged as a central degree of freedom for modulating Josephson transport[2,6,7,12]. In layered superconductors such as 2H-$NbSe_2$, which is known to exhibit strong Ising superconductivity and enhanced in-plane field robustness in the atomic limit[13,14], twist-angle engineering is widely understood to influence Cooper-pair tunneling through momentum-space matching of the superconducting electrodes[2,6,7,12]. Because the relative alignment of Fermi surfaces depends on $\theta_{twist}$, the Josephson critical current is expected to exhibit characteristic angular periodicity, with enhanced coupling near crystallographically equivalent orientations ($\theta_{\text{twist}}$ = 0°, 60°, 120°, …) and suppression at intermediate angles[2,6,7,12]. This behavior has been interpreted as evidence that twist-controlled momentum conservation governs interlayer superconducting transport in vdW JJs.

More recently, vdW Josephson systems have also been explored in the context of the Josephson diode effect (JDE), where the critical current becomes asymmetric under magnetic field reversal, $I_c^+ \neq |I_c^-|$[4,15,16]. The diode efficiency, $\eta = \frac{I_c^+ - |I_c^-|}{I_c^+ + |I_c^-|}$, is often used to quantify this nonreciprocity. Several theoretical and experimental studies have suggested that symmetry breaking associated with twist angle or stacking configuration may provide a route to engineer nonreciprocal superconducting transport[3-8]. However, a general and reproducible connection between $\theta_{\text{twist}}$ and JDE remains unresolved, partly because interfacial imperfections[17-20] can strongly modify Josephson transport even in nominally well-aligned vdW structures.

In practice, achieving uniform and atomically clean interfaces in vdW JJs remains challenging, and variations in interface transparency, contamination, or local structural disorder[17-19] can significantly alter both the magnitude and symmetry of the supercurrent[20]. These extrinsic effects can obscure intrinsic twist-angle-dependent behavior and complicate the interpretation of reported diode-like responses in similar systems[3-8].

In this work, we systematically investigate $NbSe_2$/$NbSe_2$ vertical Josephson junctions fabricated with controlled rotational alignment and high interfacial cleanliness. Using high-resolution transmission electron microscopy (HR-TEM), we directly determine $\theta_{\text{twist}}$ and assess interface structure at the atomic scale. We find that while the Josephson coupling strength, quantified by the average critical current, defined as $I_{avg} = \frac{I_c^+ + |I_c^-|}{2}$, and $I_{\text{avg}}R_{\text{n}}$ product, exhibits a clear dependence on $\theta_{\text{twist}}$ consistent with momentum-space matching[6,7], the supercurrent nonreciprocity remains negligibly small across all devices and shows no systematic twist-angle dependence. Notably, weak diode-like signals emerge only in junctions with reduced transparency[2,6,7,12] and enhanced interfacial disorder[20]. Deliberate degradation of the interface further amplifies this apparent nonreciprocity, despite the absence of any change in twist angle or evidence of a distinct superconducting state. Across all devices, $\eta$ remains small and $\theta_{twist}$-independent, even in junctions with widely varying Josephson coupling strength. These results establish a clear decoupling between Josephson coupling and supercurrent nonreciprocity in twisted $NbSe_2$/$NbSe_2$ junctions. They further demonstrate that interface quality[17-20], rather than twist-angle-controlled momentum matching[2,6,7,12], dominates the observed diode response. Our findings provide a critical benchmark for interpreting previously reported Josephson diode effects in vdW superconducting systems[3-8] and highlight the necessity of distinguishing intrinsic symmetry-breaking physics from extrinsic interface-induced artifacts.

## 2. EXPERIMENTAL SECTION

For this study, vdW $NbSe_2$/$NbSe_2$ vertical junctions are fabricated using two $NbSe_2$ flakes mechanically exfoliated onto a $SiO_2$/Si substrate (Figures 1a,b). The $NbSe_2$ flakes are obtained by exfoliating bulk crystals purchased from HQ Graphene. To minimize tape residue and improve flake quality, the crystals are first exfoliated onto polydimethylsiloxane (PDMS) stamps and then transferred onto a $SiO_2$(300 nm)/Si substrate[4,9-11]. Flakes with thicknesses ranging from 20 to 55 nm and well-defined edges are selected, and their geometric shapes are used to approximately control the twist angle between layers. The two flakes were then stacked onto pre-patterned Ti/Au electrodes using a dry transfer technique[4,9-11]. During twist stacking, the long, straight edges of the flakes were used as a reference for the crystallographic *a*-axis, following established crystallographic alignment approaches for vdW heterostructures[22]. The twist angle is initially controlled approximately by rotating the stage during alignment. For more quantitative determination of the relative crystal orientation, HR-TEM characterization is employed[2]. Detailed schematics of the fabrication process and HR-TEM analysis are provided in the Supplementary Information. To reduce oxidation and contamination at the flake surfaces and interfaces, the stacking process was performed in a nitrogen-filled glove box with $H_2O$ and $O_2$ levels maintained below 0.5 ppm[24].

The current-voltage $I$–$V$ curves of the fabricated $NbSe_2$/$NbSe_2$ vertical JJs are measured in a Quantum Design Physical Property Measurement System over a temperature $T$ range from 300 K down to 2 K. Four-probe $I$–$V$ measurements are performed using a Keithley 6221 current source and a Keithley 2182 A nanovoltmeter. All working junctions exhibit Ohmic $I$–$V$ characteristics in the normal state, with resistance values in the range of 40–60 Ω at 300 K. A superconducting transition of such junctions is observed at approximately 6 K. The critical current $I_c$ is defined at the point where $V(I) \approx 1$ μV, following conventional Josephson-junction transport analysis[25]. We note that $\theta$ denotes the angle between the applied $\mu_0 H_{\parallel}$ and one edge of the Si wafer, which is intentionally cut into a rectangular shape to provide a well-defined reference direction. The exfoliated layer stack is aligned with this edge. Field-dependent diode signals are measured by applying $\mu_0 H_{\parallel}$ along a specified $\theta$ direction. Angle-dependent diode signals are obtained by rotating the sample with respect to a fixed superconducting electromagnet while maintaining a constant $\mu_0 H_{\parallel}$.

To characterize the thicknesses of the top and bottom $NbSe_2$ layers, atomic force microscopy (AFM) measurements are performed on the fabricated vdW vertical JJs. The AFM images and corresponding height profiles are analyzed in conjunction with optical micrographs of the junctions.

## 3. RESULTS AND DISCUSSION

**3.1. Twist and interface effects on electrical transport properties of vdW JJs.** We first investigate the twist-angle-dependent transport properties of $NbSe_2/NbSe_2$ Josephson junctions by measuring their *I*–*V* curves at $T$ = 2 K (Figure 1c). In the case of Junction A with $\theta_{twist} \approx 0^o$ (Figure 1c), the critical current $I_c$ is determined to be 1.12 mA, and Junction A exhibits clear hysteresis, indicative of underdamped Josephson dynamics. To analyze the hysteretic behavior, the junction capacitance is estimated using a parallel-plate approximation $C = \varepsilon_o \varepsilon_r A/d$. With a junction area $A$ of 17.01 μm² and an assumed interlayer spacing $d$ of 0.6 nm, the effective capacitance is $C_{eff}$ = 0.25 pF. Within the resistively and capacitively shunted junction (RCSJ) model[25-29], the McCumber parameter $\beta = (2eI_cR_n^2C)/\hbar$ is calculated using the extracted $C_{eff}$ = 0.25 pF, the normal-state resistance $R_n$ = 5.21 Ω obtained from the differential resistance in the quasiparticle regime, and $\hbar = 1.055 \times 10^{-34}$ J·s is the reduced Planck constant. The resulting value $\beta$ = 7.1 confirms that Junction A operates in the underdamped regime and therefore a hysteretic *I*–*V* characteristic is expected[25-29]. In contrast, Junction D with $\theta_{twist} \approx 30^o$ exhibits a much lower $\beta$ value (= 0.14), indicating overdamped, non-hysteretic behavior, consistent with recent reports of overdamped vdW Josephson junctions engineered through geometrical and interfacial control[27-29]. To further evaluate possible thermal contributions, we consider the dissipated power parameter $I_c^2R_n$, representing Joule heating at the switching point. We observe that hysteresis tends to decrease with reduced dissipated power, suggesting that both intrinsic Josephson dynamics (as predicted by the RCSJ model) and self-heating effects contribute to the switching behavior. This implies that the observed electrical response is also sensitive to the quality and transparency of the junction interface[27-29].

To further understand the twist-angle-dependent transport characteristics, we compare the experimental results with the Ambegaokar–Baratoff (AB) relation[30], which describes BCS-type tunneling in S/I/S junctions and serves as a benchmark for evaluating phase-coherent coupling strength, $I_cR_n = \frac{\pi\Delta(T)}{2e}\tanh\left[\frac{\Delta(T)}{2k_BT}\right]$, where $\Delta(T)$ is the superconducting energy gap for a given $T$, $k_B$ is the Boltzmann constant[25-29]. At $T$ = 0 K, this expression reduces to $I_cR_n =$ $\frac{\pi\Delta(0)}{2e}$. Using the effective average gap $\Delta_{eff}(0)$ = 1.1–1.3 meV for 2H-$NbSe_2$, which exhibits multiband superconductivity[31], we obtain a zero-temperature AB reference value of $I_cR_n$ = 1.7–2.0 mV. Comparing the experimental results with this benchmark, we find that the junctions A–C exhibit the $I_cR_n$ values of 4.6–6.2 mV at 2 K, exceeding the AB reference range. Such enhancement is often associated with highly transparent Josephson interfaces and transport approaching the clean short-junction or ballistic limit, where the conventional SIS tunneling description is no longer sufficient[25-29]. In contrast, Junction D shows a significantly reduced value ($I_cR_n$ = 0.26 mV at 2 K), consistent with suppressed and weak coupling.

An analysis based on the Josephson coupling energy, $E_J = \frac{\hbar I_c}{2e}$, which characterizes the strength of phase coupling between two superconductors in a Josephson junction, further supports this conclusion when compared with the thermal energy $k_BT$. Since Junctions A–C exhibit much larger $I_c$, they correspond to $\frac{E_J}{k_BT} \sim 10^4$, reinforcing the argument of strong phase-coherent coupling[32,33]. In contrast, Junction D, with its much smaller $I_c$, still satisfies $\frac{E_J}{k_BT} \sim 10^2$, but with a significantly reduced ratio, indicating comparatively weaker phase coupling.

**3.2. Dependences of JDE on magnetic field strength.** To elucidate the intrinsic or extrinsic origin of JDE in twisted $NbSe_2$/$NbSe_2$ vertical junctions, we systematically investigate the evolution of the differential resistance $dV/dI$ as a function of $I$ and $\mu_0H_{\parallel}$. Figure 2a presents the $dV/dI(I, \mu_0H_{\parallel})$ contour map of Junction A measured over the $\mu_0H_{\parallel}$ range of −3 to +3 T. The blue regions correspond to the zero-resistance superconducting state. The superconducting phase boundary exhibits nearly perfect symmetry with respect to both current and magnetic-field polarity, indicating negligible non-reciprocal supercurrent transport and the absence of a measurable JDE. Qualitatively similar behavior is observed for Junctions B and C (Figures 2b,c). In all three junctions, the superconducting region gradually decreases with increasing field magnitude and follows an approximately symmetric dependence that can be described by a broadened sinc-like or parabolic envelope. In contrast, Junction D exhibits a substantially stronger suppression of superconductivity under $\mu_0 H_{\parallel}$, accompanied by an irregular evolution of the superconducting phase boundary in the low-field regime ($|\mu_0 H_{\parallel}| < 0.3$ T). This behavior differs markedly from that of Junctions A–C and suggests a reduced Josephson coupling strength associated with a less transparent interface. Such an interpretation is consistent with the significantly lower $I_c R_n$ product and $E_J$ inferred for Junction D, indicating weakened phase coherence across the junction[25-29]. The enhanced magnetic-field sensitivity of Junction D and irregular low-field response of Junction D are attributed to a more weakly phase-coupled Josephson junction, likely arising from a combination of largest Fermi-surface misalignment (i.e., approaching $\theta_{twist} = 30°$)[2,6,7,12] and increased interfacial inhomogeneity[17-20]. We further note that the irregular evolution of the superconducting phase boundary in this device is influenced by thermal rounding of the current–voltage characteristics near the switching transition, which becomes more pronounced in the weak-coupling regime[25,26].

The magnetic-field dependence of the diode efficiency $\eta(\mu_0 H_{\parallel})$ provides additional insight into the origin of non-reciprocal transport (Figure 2e). Junction A exhibits negligible $\eta$ over the entire field range, within the experimental resolution of our measurement system, confirming the absence of measurable non-reciprocal supercurrent. For Junctions B and C, $\eta$ remains near zero at low magnetic fields and increases only slightly at higher fields, remaining below 8% throughout the measured range. In contrast, Junction D displays a finite and highly irregular $\eta(\mu_0 H_{\parallel})$ in the low-field regime, consistent with its anomalous magnetic-field response. The correlation between the irregular magnetic-field response and the enhanced apparent diode signal (up to 11%) in Junction D suggests that the observed non-reciprocity originates primarily from extrinsic interface-related effects rather than intrinsic $\theta_{twist}$-dependent superconducting physics[6,7,12]. Taken together, these results demonstrate that all investigated junctions exhibit negligible diode efficiencies irrespective of twist angle, highlighting that local interface inhomogeneity can produce apparent non-reciprocal transport signatures that may obscure the intrinsic behavior of twisted $NbSe_2/NbSe_2$ Josephson junctions[4,15,16].

**3.3. Dependences of JDE on magnetic field angle.** To further examine the origin of JDE in twisted $NbSe_2/NbSe_2$ vertical junctions, we investigate the angular dependence of *dV/dI* map and $\eta$ under a fixed in-plane magnetic field of $\mu_0 H_{\parallel}$ = 0.07 T. This field is selected because magnetic-field-induced transport features are most clearly resolved in this regime while the junction's superconductivity is retained for all the devices studied. Figures 3a–d show the evolution of the *dV/dI* maps as the in-plane field angle $\gamma$ varies in 10° increments. Junctions A–C exhibit nearly $\gamma$-independent superconducting phase boundaries, with the critical-current branches $I_c^{\pm}$ remaining essentially unchanged throughout the full angular range. The absence of any discernible modulation indicates that the superconducting transport in these junctions is largely insensitive to the orientation of the applied $\mu_0 H_{\parallel}$. Such behavior is consistent with the highly symmetric field response observed in the magnetic-field-strength-dependent measurements (Figure 2) and suggests that neither $\theta_{\text{twist}}$ nor the field orientation induces measurable non-reciprocal supercurrent transport in these devices. In contrast, Junction D displays a markedly different response. The *dV*/*dI* map reveals a pronounced zigzag-like modulation of the superconducting phase boundary with $\gamma$, indicating enhanced sensitivity to the magnetic-field orientation. This behavior is accompanied by substantial variations in the extracted diode efficiency, implying that the transport properties of Junction D are strongly influenced by local interfacial characteristics[17-20]. The corresponding angular dependence of the diode efficiency $\eta(\gamma)$ is summarized in Figure 3e. Junction A exhibits a finite but extremely small $\eta$ over the entire angular range, remaining close to the experimental resolution limit. Junctions B and C show $\eta$ values that are nearly zero and independent of $\gamma$, further confirming the absence of an intrinsic JDE. By contrast, Junction D exhibits a significantly larger apparent diode response, reaching a maximum $\eta$ of approximately 11%. Notably, however, the angular dependence does not follow a well-defined symmetry or 60° periodicity expected from an intrinsic twist-angle-related mechanism[2,6,7,12]. Instead, $\eta(\gamma)$ varies irregularly with field angle and remains finite across the entire angular range. The absence of systematic angular modulation in Junctions A–C, together with the irregular behavior observed in Junction D, suggests that the measured non-reciprocal response is not governed by $\theta_{\mathrm{twist}}$-dependent superconducting physics. Rather, the enhanced $\eta$ observed in Junction D is more consistent with extrinsic effects arising from interface inhomogeneity[17-20] and reduced junction transparency[2,6,7,12].

**3.4. Temperature evolution of Josephson supercurrent non-reciprocity.** To further clarify the origin of the apparent non-reciprocal supercurrent response, we investigate the temperature dependence of *dV/dI* and $\eta$ in the $NbSe_2$/$NbSe_2$ vertical Josephson junctions. Figure 4a shows the *dV/dI* map of Junction A as a function of *I* and *T*, measured under a fixed $\mu_0 H_{\parallel} = 0.07$ T. The dark-blue region corresponds to the zero-resistance superconducting state of the junction. As *T* increases, the superconducting phase boundary gradually shrinks and ultimately disappears above the junction's superconducting transition temperature, reflecting the progressive suppression of Josephson coupling by thermal fluctuations[25,26]. Junctions B-D show similar *T* dependences. In all devices, the superconducting phase boundary decreases monotonically with increasing *T* and vanishes near their respective transition temperature. Despite the markedly different magnetic-field-strength and magnetic-field-angle responses observed for Junction D in Sections 3.2 and 3.3, its *T* evolution closely resembles that of the strongly coupled junctions. No anomalous *T*-dependent features unique to Junction D are observed. This similarity indicates that all investigated devices share the same underlying superconducting transport mechanism and that the distinct behavior of Junction D originates primarily from interfacial properties[17-20] rather than from a fundamentally different superconducting state[7].

The temperature dependence of the diode efficiency $\eta(T)$ is summarized in Figure 4e. For the strongly coupled Junctions A–C, $\eta$ remains close to zero throughout the entire measured $T$ range, confirming the absence of measurable non-reciprocal supercurrent transport. By contrast, Junction D exhibits a finite $\eta$ that gradually decreases with increasing $T$ and approaches zero near the transition temperature. The suppression of $\eta(T)$ closely follows the reduction of the superconducting switching current (Figure 1d), indicating that the apparent non-reciprocal response weakens as superconductivity is thermally suppressed. Importantly, $\eta$ in Junction D does not reveal any additional characteristic $T$ dependence[7] or anomalous behavior that would suggest an intrinsic diode mechanism associated with the twist angle. Therefore, although Junction D exhibits a larger apparent diode response than the other devices, its overall $T$ evolution remains consistent with conventional Josephson behavior[25,26].

**3.5. Twist-angle dependent Josephson supercurrents and twist-angle independent non-reciprocity.** Figures 5a and 5b provide a direct comparison between the $\theta_{\text{twist}}$ dependences of Josephson coupling, quantified by the characteristic voltage $V_c$, and supercurrent non-reciprocity, characterized by $\eta$. Here, $V_c$ is defined as $I_{avg}R_n$, while $\eta$ represents the diode efficiency attained under a fixed $\mu_0H_{\parallel} = 0.07$ T for each junction. As shown in Figure 5a, the Josephson coupling strength exhibits a pronounced dependence on $\theta_{\text{twist}}$. Junctions with $\theta_{\text{twist}}$ close to the crystallographically equivalent orientations (～$0^{\circ}$ and ～$60^{\circ}$) show substantially larger $V_c$ than those with intermediate $\theta_{\text{twist}}$. This behavior is consistent with the expected momentum-conserving Josephson tunneling picture, in which the overlap between the Fermi surfaces of the two $NbSe_2$ electrodes depends sensitively on their relative crystallographic orientation[2,6,7,12]. Near symmetry-equivalent stacking configurations, the overlap of electronic states is maximized, resulting in enhanced Cooper-pair transmission and stronger Josephson coupling[2,6,7,12]. In contrast, larger momentum mismatch at intermediate twist angles suppresses coherent tunneling and reduces the critical current. Within the *c*-axis twist-junction framework[12], the angular dependence of the Josephson coupling can be expressed as $I_{avg}R_n(\theta_{twist}) \propto \int d\boldsymbol{k} T_{\boldsymbol{k}}^2 \Delta(\boldsymbol{k})\Delta(R_{\theta_{twist}}\boldsymbol{k})$, where the integral is taken over the Fermi surface, $R_{\theta_{twist}}$ rotates momentum by the twist angle $\theta_{twist}$, $T_{\boldsymbol{k}}$ is the tunneling matrix element, $\Delta(\boldsymbol{k})$ is the superconducting gap. For 2H-$NbSe_2$, $\Delta(\boldsymbol{k})$ inherits a six-fold symmetry from the underlying hexagonal lattice[13,14]. As a result, $I_{avg}R_n(\theta_{twist})$ exhibits periodic modulations governed by crystal symmetry and momentum-selective tunneling, approximately captured by a 60◦-periodic form such as $I_{avg}R_n(\theta_{twist}) \propto |\cos(3\theta_{twist})|$ (dashed line in Figure 5a). The observed trend therefore confirms that the Josephson coupling strength in clean $NbSe_2$/$NbSe_2$ vertical junctions remains strongly governed by twist-angle-controlled momentum-space alignment[2,6,7,12].

In striking contrast, the Josephson supercurrent non-reciprocity exhibits no systematic dependence on $\theta_{twist}$ (Figure 5b). For all investigated junctions, $\eta$ remains small and does not follow the trend observed for $V_c$. Although Junction D exhibits a somewhat larger apparent diode efficiency than the other devices, this enhancement occurs despite its significantly weaker Josephson coupling and does not correlate with any specific $\theta_{twist}$. Furthermore, as demonstrated in Sections 3.2–3.4, the enhanced $\eta$ of Junction D is accompanied by irregular magnetic-field and angular responses, suggesting that it originates from extrinsic interface-related effects[17-20] rather than an intrinsic $\theta_{twist}$-induced mechanism[2,6,7,12]. Consequently, the overall distribution of $\eta$ across the measured $\theta_{twist}$ range indicates that the non-reciprocal supercurrent response is essentially independent of $\theta_{twist}$.

The contrasting behaviors of $V_c$ and $\eta$ provide an important insight into the physics of twisted $NbSe_2$/$NbSe_2$ vertical Josephson junctions. While twist-angle engineering effectively modulates the Josephson coupling strength through momentum-space matching of the superconducting electrodes, it does not generate a corresponding enhancement of non-reciprocal supercurrent transport when the interfaces are chemically clean and structurally uniform. Instead, any residual diode-like response appears to be dominated by local interface inhomogeneity, transparency variations, or other extrinsic factors[17-20]. These findings establish a clear distinction between the mechanisms governing Josephson coupling and supercurrent non-reciprocity and indicate that large Josephson diode effects reported in similar vdW junction systems[3-8] should be carefully evaluated in terms of interface quality[17-20] in addition to twist-angle considerations[2,6,7,12].

**3.6. Extrinsic enhancement of apparent Josephson diode effects.** To further clarify the origin of the apparent non-reciprocal response, we focus on the weak-coupling regime, where Josephson transport is strongly suppressed and becomes highly sensitive to local interface conditions. In particular, junctions with twist angles near 30° exhibit the smallest critical currents and the lowest $I_cR_n$ products, consistent with reduced momentum-space overlap and weakened phase coherence between the $NbSe_2$ electrodes[2,6,7,12]. In this regime, we find that the measured diode efficiency can be substantially enhanced when the junction interface is intentionally degraded. Specifically, when the post-stacking bakeout step is omitted[4,9-11], the junctions exhibit a more pronounced apparent non-reciprocal response within a restricted magnetic-field window (Figures. 6a-c). In these devices, the extracted diode efficiency $\eta$ can reach values as high as ~30%. Importantly, this enhancement does not coincide with any systematic change in $\theta_{twist}$ or with signatures of a distinct superconducting state.

Instead, the large apparent $\eta$ is accompanied by strongly irregular and non-monotonic magnetic-field dependence of the critical current, as well as enhanced sensitivity to small variations in field strength. Such behavior indicates a breakdown of uniform Josephson coupling across the junction area, consistent with spatially inhomogeneous interface transparency[17-20] and reduced phase stiffness[25,26]. In this situation, current distribution becomes highly non-uniform, and local switching events can be influenced by microscopic disorder[20] and localized heating near the critical current. These effects can produce different effective switching thresholds for positive and negative bias currents, thereby mimicking a diode-like response.

The observation that the strongest apparent non-reciprocity arises in the weakest and most disordered junctions provides straightforward evidence that the enhanced Josephson diode effect in these devices is not intrinsic in origin. Rather, it emerges from extrinsic factors associated with interface inhomogeneity[17-20], including spatial variations in transparency[2,6,7,12], local suppression of superconductivity, and non-uniform current flow. This interpretation is fully consistent with the $\theta_{\mathrm{twist}}$-independent behavior of $\eta$ discussed in Section 3.5, and it reinforces the conclusion that crystallographic misalignment does not play a primary role in generating non-reciprocal supercurrent transport in clean $NbSe_2$/$NbSe_2$ vertical Josephson junctions. Taken together, these results demonstrate that apparent diode effects can be artificially amplified in the weak-coupling regime due to interface disorder[17-20], rather than arising from an intrinsic symmetry-breaking mechanism[2,6,7,12] linked to $\theta_{\mathrm{twist}}$.

## 4. CONCLUSIONS

In summary, we have systematically investigated the role of $\theta_{\mathrm{twist}}$ in determining both the Josephson coupling strength and the non-reciprocal supercurrent response in $NbSe_2$/$NbSe_2$ vertical JJs fabricated via a dry transfer technique with improved interface cleanliness. While the critical current and Josephson coupling energy exhibit a clear dependence on $\theta_{\mathrm{twist}}$, consistent with variations in momentum-space overlap between the superconducting layers[2,6,7,12], the non-reciprocal supercurrent response remains negligibly small across all measured devices and shows no systematic dependence on $\theta_{twist}$. Despite observing a wide range of junction transparencies, from strongly coupled junctions with anomalously high $I_cR_n$ values exceeding the AB limit[30] to weakly coupled junctions approaching the conventional tunneling regime, the measured $\eta$ remains near zero in all cases. Only a weak and non-universal enhancement of asymmetry is observed in weakly coupled junctions within a limited magnetic-field window, suggesting that any apparent diode-like behavior is not intrinsic to $\theta_{twist}$ but is instead highly sensitive to interfacial inhomogeneity and junction quality[17-20]. These results demonstrate that twist-angle engineering[2,6,7,12] alone is not sufficient to induce robust non-reciprocal superconducting transport in $NbSe_2$-based vertical Josephson junctions when high-quality, chemically clean interfaces are realized. Instead, our findings highlight interface transparency[2,6,7,12] and microscopic disorder[17-20] as the dominant factors governing any residual diode response. This provides a revised framework for interpreting previously reported large JDEs in similar van der Waals systems[6-8] and establishes an important benchmark for future studies aiming to exploit layered superconductors for controlled non-reciprocal superconducting electronics.

## ASSOCIATED CONTENT

**Supporting Information.** The Supporting Information is available free of charge at https://pubs.acs.org/doi/XXX. Sections S1–S4 include the following: fabrication process of the van der Waals Josephson junctions; optical microscopy and atomic force microscopy characterization of the fabricated devices; high-resolution transmission electron microscopy characterization used to determine the relative crystal orientation and twist angle; and Josephson properties of Junctions E and F with nominal twist angles of 15° and 40°, respectively.

## ACKNOWLEDGMENTS

**Funding.** This work was supported by the Max Planck Partner Group 2023, the POSCO Science Fellowship 2024, the National Research Foundation (NRF) of Korea (Grant No. 2020R1A5A1016518, RS-2026-25488513), and the European Union (ERC Advanced Grant SUPERMINT, project number 101054860). S.-G. L. gratefully acknowledges Tobias Fäth for valuable guidance on the use of a 2D transfer stage at MPI-MSP. **Author Contributions.** K.-R.J. and S.S.P.P. conceived and designed the experiments. S.-G.L. fabricated the van der Waals Josephson junctions under the guidance of J.-K.K. and K.-R.J. S.-G.L. and K.-N.J. carried out electrical transport measurements, atomic force microscopy characterization, and transmission electron microscopy–based microstructural and compositional analyses under the supervision of K.-R.J. J.-C.J. and J.Y. contributed to the electrical transport measurements. S.-G.L. analyzed the transport data with guidance from K.-R.J. K.-R.J., S.-G.L., and S.S.P.P. wrote the manuscript with input from all coauthors. **Competing Interests.** The authors declare no competing interests. **Data and Materials Availability.** All data required to evaluate the conclusions of this study are included in the article and/or the Supporting Information.

## FIGURE CAPTIONS

**Figure 1. Structural characterization and electrical transport properties of $NbSe_2$/$NbSe_2$ van der Waals Josephson junctions.** (a) Optical micrographs of Junctions A–D with nominal twist angles of 0°, 45°, 60°, and 30°, respectively. (b) Cross-sectional high-resolution transmission electron microscopy (HR-TEM) images of the $NbSe_2$/$NbSe_2$ interfaces for Junctions A–D. (c) Current–voltage *I*–*V* characteristics of Junctions A–D measured at 2 K. (d) Critical current $I_c$ as a function of temperature $T$ for Junctions A–D.

**Figure 2. Magnetic-field dependence of superconducting transport and diode efficiency.** (a–c) Differential-resistance $dV/dI$ contour maps as functions of bias current $I$ and in-plane magnetic field $\mu_0 H_\parallel$ for Junctions A–C measured at 2 K over the field range $-3\ \mathrm{T} \leq \mu_0 H_\parallel \leq 3\ \mathrm{T}$. Blue regions correspond to the superconducting state with near-zero differential resistance. (d) Differential-resistance contour map for Junction D measured under the same conditions. (e) Diode efficiency $\eta$ extracted from the positive and negative critical currents ($I_c^+$, $I_c^-$) as a function of $\mu_0 H_\parallel$ for Junctions A–D.

**Figure 3. Angular dependence of superconducting transport and diode efficiency under an in-plane magnetic field.** (a–c) Differential-resistance $dV/dI$ contour maps as functions of bias current $I$ and in-plane magnetic-field angle $\gamma$ for Junctions A–C measured at 2 K under a fixed in-plane field of $\mu_0 H_\parallel = 0.07$ T. (d) Differential-resistance contour map for Junction D measured under the same conditions. (e) Diode efficiency $\eta$ as a function of $\gamma$ for Junctions A–D, extracted from the corresponding critical currents ($I_c^+$, $I_c^-$).

**Figure 4. Temperature dependence of superconducting transport and diode efficiency.** (a–d) Differential-resistance $dV/dI$ contour maps as functions of bias current $I$ and temperature $T$ for Junctions A–D measured under a fixed in-plane magnetic field of $\mu_0 H_\parallel = 0.07$ T. (e) Diode efficiency $\eta$ as a function of $T$ for Junctions A–D, extracted from the corresponding critical currents ($I_c^+$, $I_c^-$).

**Figure 5. Twist-angle dependence of Josephson coupling and supercurrent non-reciprocity.** (a) Characteristic voltage $V_c$ (= $I_{avg}R_n$) as a function of twist angle $\theta_{twist}$ for Junctions A–F. Note that detailed data for Junctions E and F are presented in Supplementary Information. The dashed line in (a) exhibits a periodic modulation of $I_{avg}R_n(\theta_{twist}) \propto |\cos(3\theta_{twist})|$, reflecting the crystal symmetry and momentum-selective tunneling, and is approximately captured by a 60∘-periodic form. (b) Diode efficiency $\eta$ as a function of $\theta_{twist}$ for Junctions A–F, measured under a fixed in-plane magnetic field of $\mu_0 H_\parallel = 0.07$ T at 2 K. The data reveal a pronounced twist-angle dependence of the Josephson coupling, whereas the supercurrent non-reciprocity remains small and exhibits no systematic dependence on $\theta_{twist}$. In both (a) and (b), closed (open) symbols indicate nominal (estimated) twist angles; further details are provided in the Supplementary Information.

**Figure 6. Interface-degradation-induced apparent Josephson diode effect.** (a,b) Voltage $V$ and differential-resistance $dV/dI$ contour maps as functions of bias current $I$ and in-plane magnetic field $\mu_0 H_{\parallel}$ for Junction R, measured at 2 K over the field range $-0.2\ \mathrm{T} \le \mu_0 H_{\parallel} \le 0.2$ T. The blue region in (b) corresponds to the superconducting state with near-zero differential resistance. (c) Diode efficiency $\eta$ extracted from the positive and negative critical currents ($I_c^+$, $I_c^-$) as a function of $\mu_0 H_{\parallel}$ for Junction R. (d–f) Corresponding measurements for Junction D under the same conditions, shown for direct comparison. Note that the degraded-interface Junction R exhibits a substantially enhanced apparent diode response and a more irregular magnetic-field dependence than Junction D, highlighting the strong influence of interface quality on the observed non-reciprocal transport.

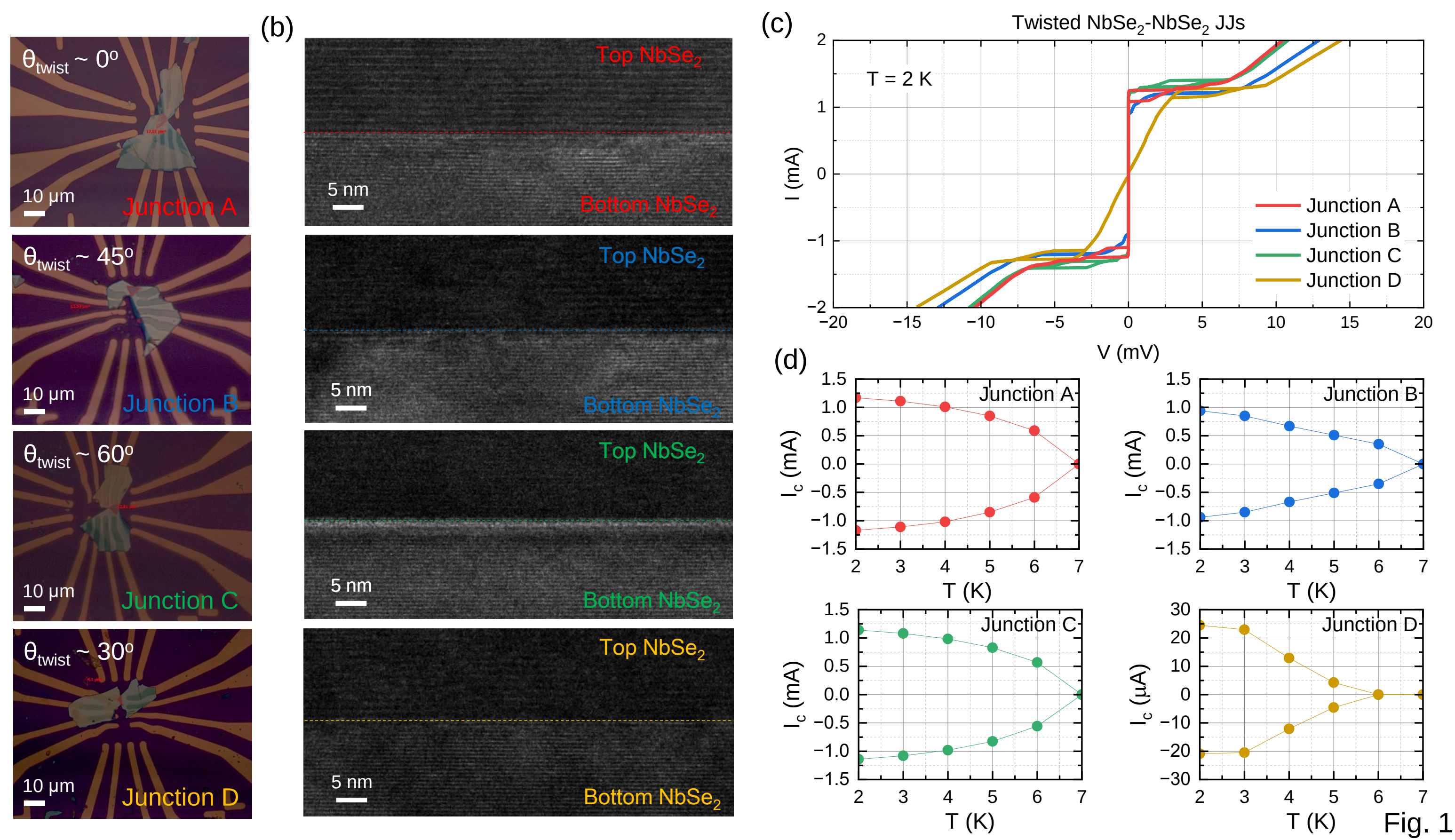
(a)
θtwist ~ 0°
10 μm
Junction A
θtwist ~ 45°
10 μm
Junction B
θtwist ~ 60°
10 μm
Junction C
θtwist ~ 30°
10 μm
Junction D
(b)
Top NbSe2
5 nm
Bottom NbSe2
Top NbSe2
5 nm
Bottom NbSe2
Top NbSe2
5 nm
Bottom NbSe2
Top NbSe2
5 nm
Bottom NbSe2
(c)
Twisted NbSe2-NbSe2 JJs
T = 2 K
I (mA)
V (mV)
Junction A
Junction B
Junction C
Junction D
(d)
Junction A
Junction B
Junction C
Junction D
Ic (mA)
Ic (μA)
T (K)

Fig. 1

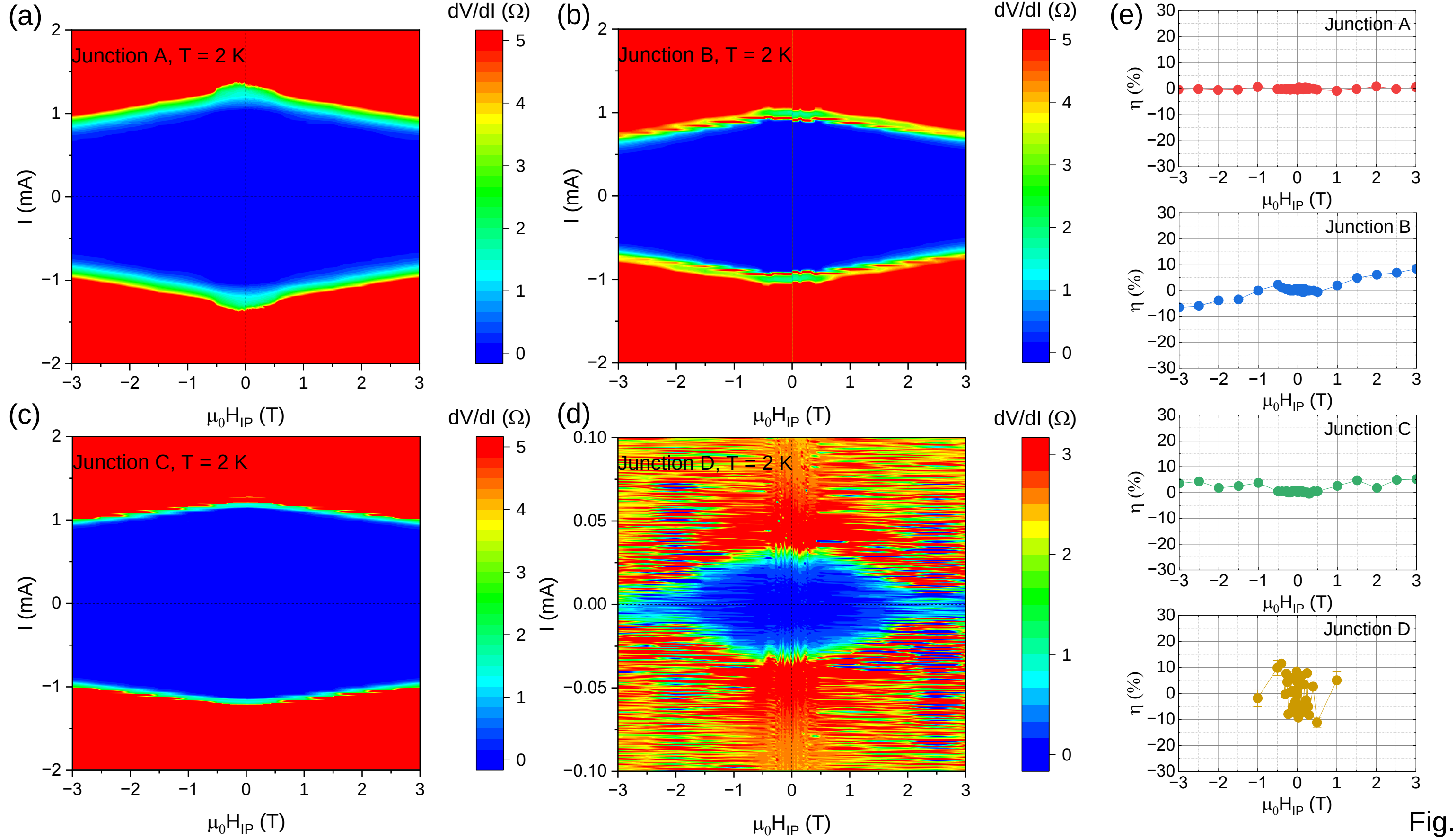

(a)
Junction A, T = 2 K
I (mA)
$\mu_0 H_{IP}$ (T)
dV/dI (Ω)
(b)
Junction B, T = 2 K
I (mA)
$\mu_0 H_{IP}$ (T)
dV/dI (Ω)
(c)
Junction C, T = 2 K
I (mA)
$\mu_0 H_{IP}$ (T)
dV/dI (Ω)
(d)
Junction D, T = 2 K
I (mA)
$\mu_0 H_{IP}$ (T)
dV/dI (Ω)
(e)
Junction A
Junction B
Junction C
Junction D
η (%)
$\mu_0 H_{IP}$ (T)

Fig. 2

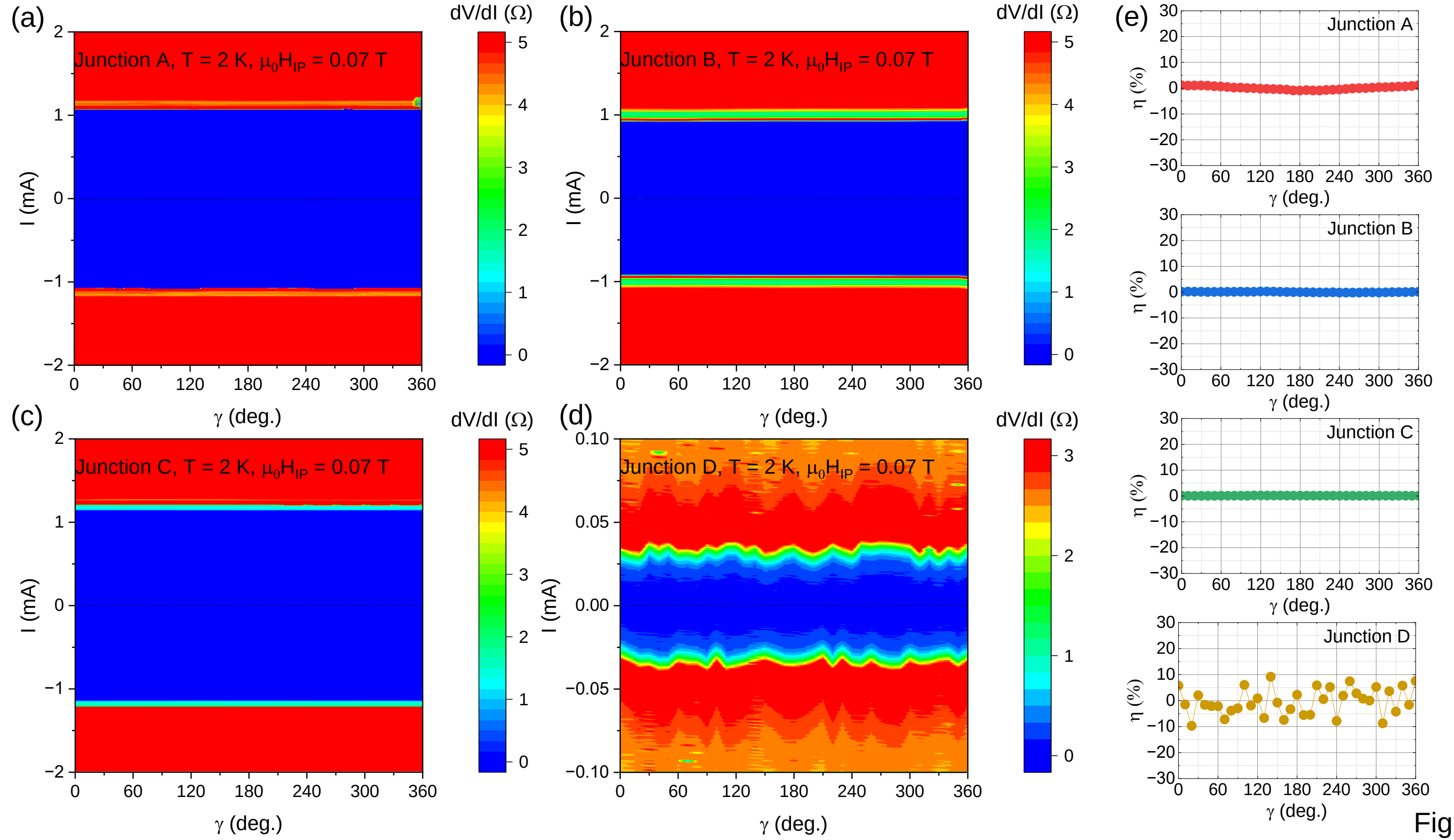
(a)
dV/dI (Ω)
Junction A, T = 2 K, $\mu_0H_{IP}$ = 0.07 T
I (mA)
γ (deg.)
(b)
dV/dI (Ω)
Junction B, T = 2 K, $\mu_0H_{IP}$ = 0.07 T
(c)
dV/dI (Ω)
Junction C, T = 2 K, $\mu_0H_{IP}$ = 0.07 T
(d)
dV/dI (Ω)
Junction D, T = 2 K, $\mu_0H_{IP}$ = 0.07 T
(e)
Junction A
Junction B
Junction C
Junction D
η (%)

Fig. 3

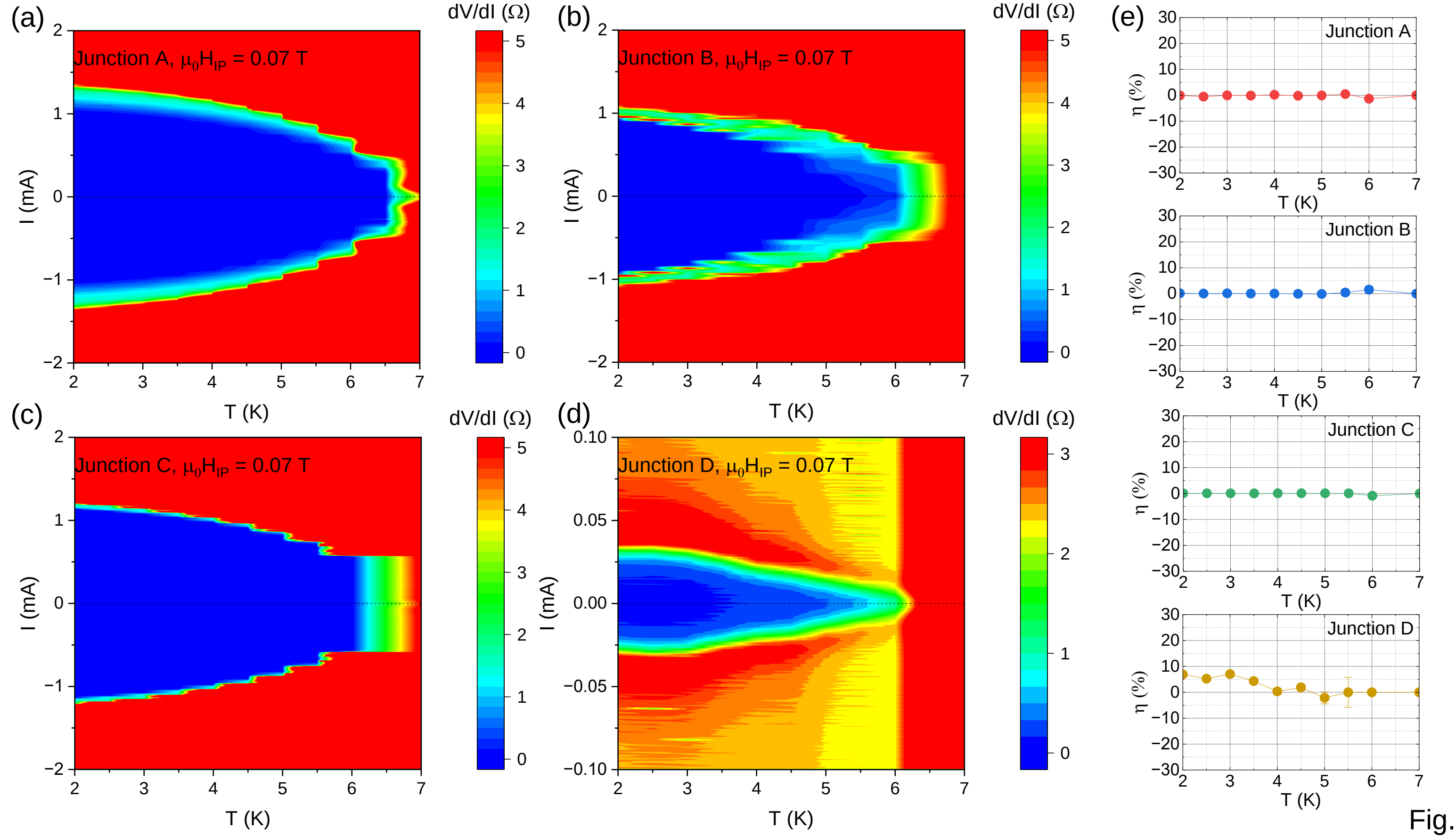

(a)
Junction A, $\mu_0H_{IP}$ = 0.07 T
dV/dI (Ω)
I (mA)
T (K)
(b)
Junction B, $\mu_0H_{IP}$ = 0.07 T
dV/dI (Ω)
I (mA)
T (K)
(c)
Junction C, $\mu_0H_{IP}$ = 0.07 T
dV/dI (Ω)
I (mA)
T (K)
(d)
Junction D, $\mu_0H_{IP}$ = 0.07 T
dV/dI (Ω)
I (mA)
T (K)
(e)
Junction A
Junction B
Junction C
Junction D
η (%)
T (K)

Fig. 4

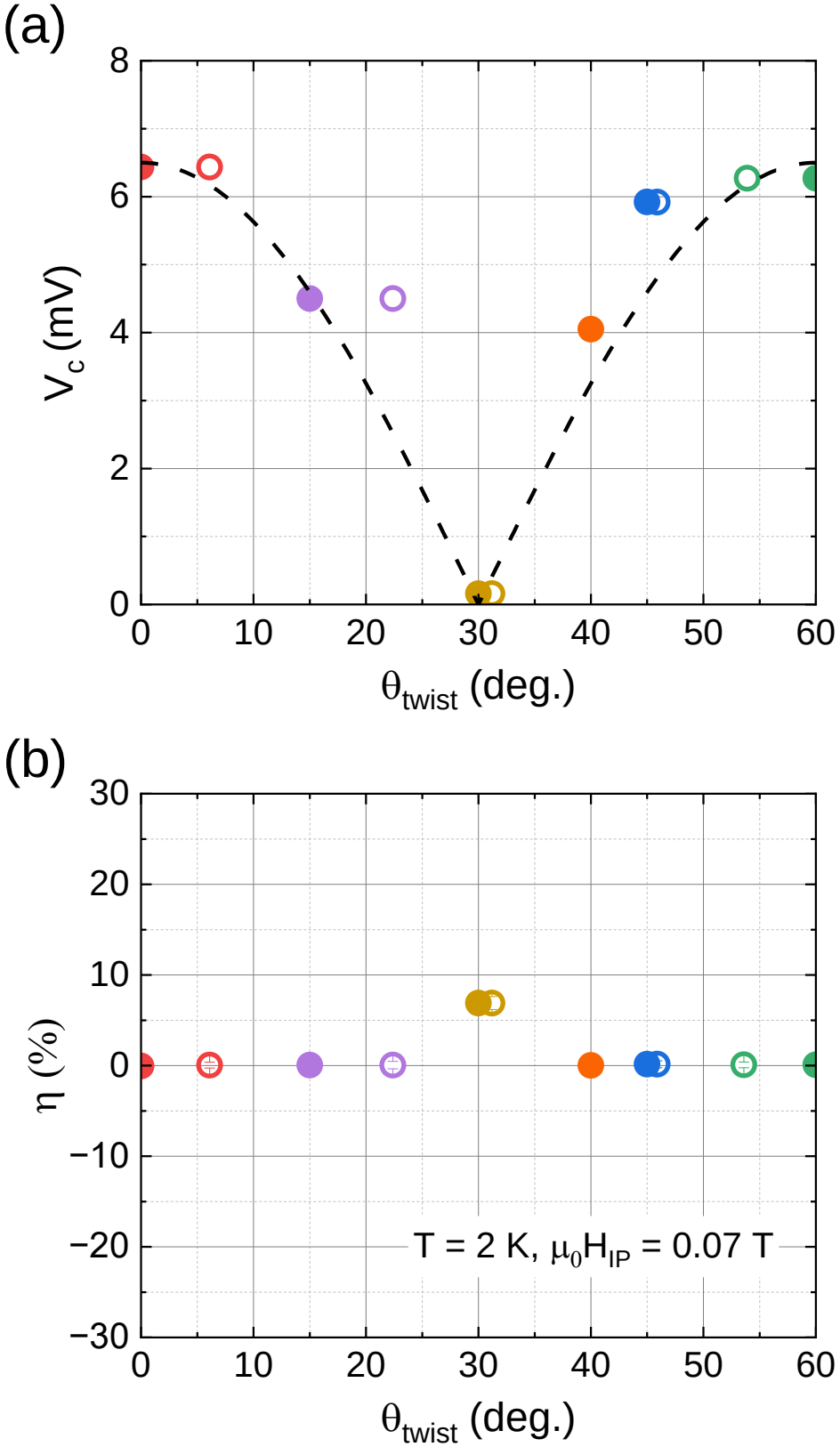
(a)
8
6
4
2
0
$V_c$ (mV)
0
10
20
30
40
50
60
$\theta_{twist}$ (deg.)
(b)
30
20
10
0
−10
−20
−30
η (%)
T = 2 K, $\mu_0H_{IP}$ = 0.07 T
0
10
20
30
40
50
60
$\theta_{twist}$ (deg.)

Fig. 5

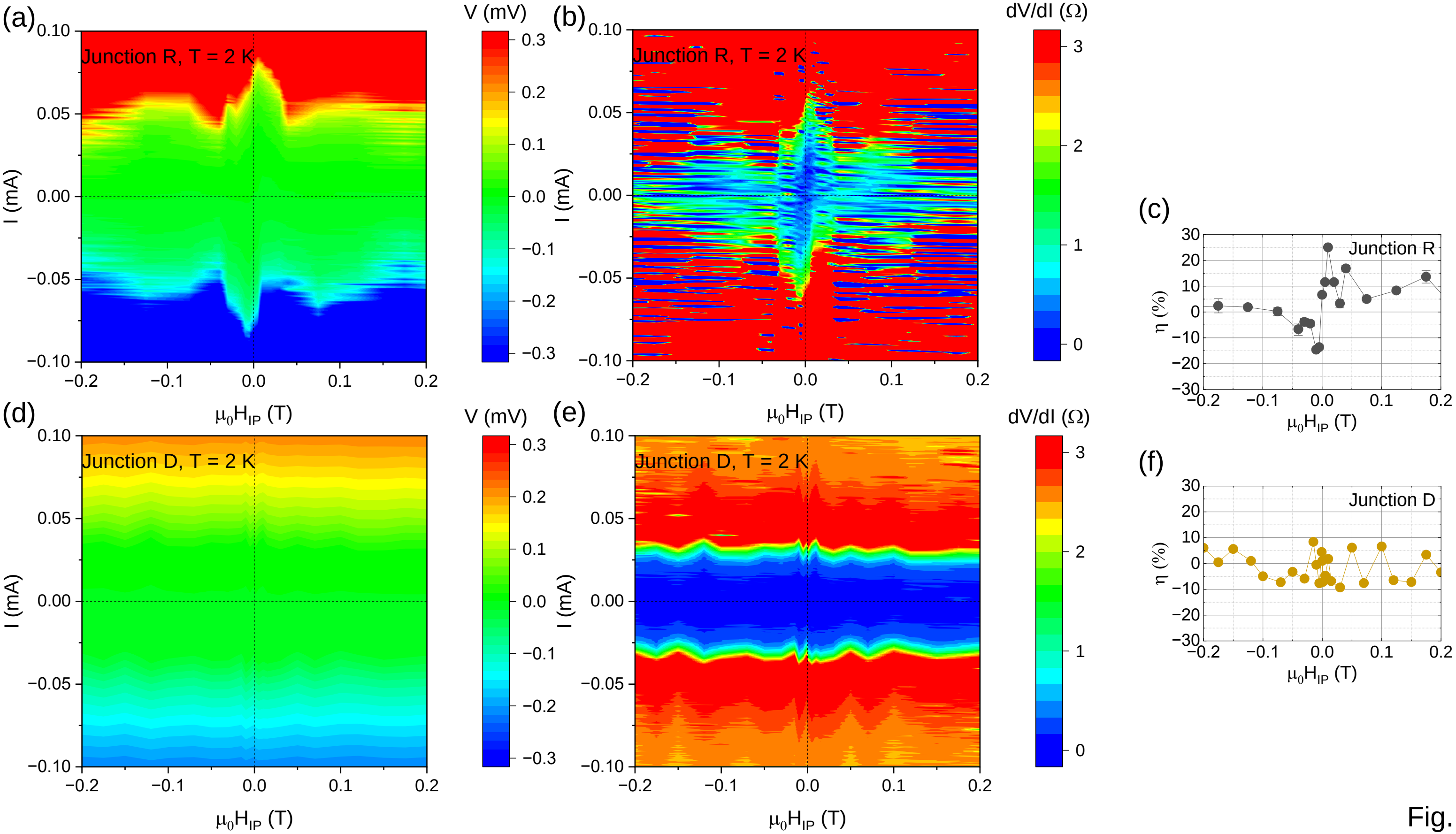

(a)
Junction R, T = 2 K
V (mV)
I (mA)
μ0HIP (T)
(b)
Junction R, T = 2 K
dV/dI (Ω)
I (mA)
μ0HIP (T)
(c)
Junction R
η (%)
μ0HIP (T)
(d)
Junction D, T = 2 K
V (mV)
I (mA)
μ0HIP (T)
(e)
Junction D, T = 2 K
dV/dI (Ω)
I (mA)
μ0HIP (T)
(f)
Junction D
η (%)
μ0HIP (T)

Fig. 6